\documentclass{article}
\usepackage{spconf}

\usepackage[english]{babel}

\usepackage{ifpdf}

\usepackage{cite} 
\usepackage{url}
\usepackage{hyperref}

\usepackage[pdftex]{graphicx}
\graphicspath{{./figures/}}
\usepackage{color}
\usepackage{pgf, tikz}
\usetikzlibrary{shapes, arrows, automata, plotmarks}
\usetikzlibrary{calc,decorations}

\usepackage[cmex10]{amsmath}
\usepackage{amsfonts, amssymb, amsthm}
\usepackage{mathrsfs}

\usepackage{algorithm,algpseudocode}
\algnewcommand{\LeftComment}[1]{\Statex \(\triangleright\) #1}

\usepackage{enumerate}
\usepackage{multirow}
\usepackage{rotating}
\usepackage{subcaption}
\input{pennColors.sty}
\usepackage{needspace}

\input{mySymbol.sty}

\def\Tr{\mathsf{T}}

\title{OPTIMAL POWER FLOW USING GRAPH NEURAL NETWORKS}
\name{Damian Owerko, Fernando Gama, and Alejandro Ribeiro\thanks{Supported by NSF CCF 1717120, ARO W911NF1710438, ARL DCIST CRA W911NF-17-2-0181, ISTC-WAS and Intel DevCloud.}}
\address{Department of Electrical and Systems Engineering, University of Pennsylvania}
\begin{document}
\ninept
\maketitle
\begin{abstract}
Optimal power flow (OPF) is one of the most important optimization problems in the energy industry. In its simplest form, OPF attempts to find the optimal power that the generators within the grid have to produce to satisfy a given demand. Optimality is measured with respect to the cost that each generator incurs in producing this power. The OPF problem is non-convex due to the sinusoidal nature of electrical generation and thus is difficult to solve. Using small angle approximations leads to a convex problem known as DC OPF, but this approximation is no longer valid when power grids are heavily loaded. Many approximate solutions have been since put forward, but these do not scale to large power networks. In this paper, we propose using graph neural networks (which are localized, scalable parametrizations of network data) trained under the imitation learning framework to approximate a given optimal solution. While the optimal solution is costly, it is only required to be computed for network states in the training set. During test time, the GNN adequately learns how to compute the OPF solution. Numerical experiments are run on the IEEE-30 and IEEE-118 test cases.
\end{abstract}
\begin{keywords}
graph neural networks, smart grids, optimal power flow, imitation learning
\end{keywords}
%


\vspace{-0.15cm}

\section{Introduction} \label{sec:intro}

Optimal power flow (OPF) is one of the most important optimization problems for the energy industry \cite{acopf}. It is used for system planning, establishing prices on day-ahead markets, and to allocate generation capacity efficiently throughout the day. Even though the problem was formulated over half a century ago, we still do not have a fast and robust technique for solving it, which would save tens of billions of dollars annually \cite{acopf}.

In its simplest form, OPF attempts to find the optimal power that the generators within the grid have to produce to satisfy a given demand. Optimality is measured with respect to the cost that each generator incurs in producing the required power. Even though this problem is at the heart of daily electricity grid operation, the sinusoidal nature of electrical generation makes it difficult to solve \cite{acopf,nonconvex,bienstock}. One alternative is to linearly approximate the problem by assuming constant bus voltage and taking small angle approximations for trigonometric terms of voltage angle difference. This approach has limitations. The small angle approximations fail for heavily-loaded networks, since differences in voltage angles become large \cite{lecture}. Nevertheless, this approximation is commonly used in industry \cite{acopf}. The exact formulation of the problem is commonly referred to as ACOPF and the approximation as DCOPF.

Incorporating AC dynamics results in a non-convex problem due to nonlinearities in the power flow equations \cite{nonconvex, molzahn} and has been proven to be NP hard \cite{bienstock, acopf}. One approach is to apply convex relaxation techniques resulting in semi-definite programs \cite{molzahn}. There have been many attempts to apply machine intelligence to the problem. The work in \cite{intelligence} offers a comprehensive review of such approaches. These include include applying evolutionary programming, evolutionary strategies, genetic algorithms, artificial neural networks, simulated annealing, fuzzy set theory, ant colony optimization and particle swarm optimization. However, none of these approaches has been shown to work on networks larger than the 30 node IEEE test case \cite{bakirtzis,todorovski,particle}.

Recently, motivated by the possibility of accurately generating large amounts of data, machine learning methods have been considered as solutions to this problem. More specifically, \cite{dobbe} propose a system that uses multiple stepwise regression \cite{sondermeijer} to imitate the output of ACOPF. Each node gathers information from a subset of nodes, although not necessarily neighboring nodes. They then use multiple stepwise regression to predict the optimal amount of power generated at each node, imitating the ACOPF solution. However, this solution is not local, since it uses information from nodes that are not adjacent in the network. The work in \cite{guha} uses a fully connected network (MLP) to imitate the output of ACOPF. Yet, MLPs are not local either, tend to overfit and have trouble scaling up. Instead, exploiting the structure of the problem is necessary for a scalable solution. An exact, albeit costly solution, can be obtained by using interior point methods to solve the ACOPF problem, which, in practice, converges to the optimal solution -- though not always and without guarantee of optimality \cite{acopf, bienstock}.

In this work, we use \emph{imitation learning} and \emph{graph neural networks} \cite{Bruna14-DeepSpectralNetworks, Defferrard17-CNNGraphs, Gama19-Architectures} to find a local and scalable solution to the OPF problem. More specifically, we adopt a parametrized GNN model, which is local and scalable by design, and we train it to \emph{imitate} the optimal solution obtained using an interior point solver \cite{pandapower}, which is centralized and does not converge for large networks. Once trained, the GNN offers an efficient computation of ACOPF. The paper is structured as follows. In Sec.~\ref{sec:OPF} we describe the OPF problem and in Sec.~\ref{sec:GNN} we introduce GNNs. In Sec.~\ref{sec:sims} we test the imitation learning framework on the IEEE-30 and IEEE-118 power system test cases \cite{matpower} and in Sec.~\ref{sec:conclusions} we draw conclusions.


\vspace{-0.2cm}

\section{Optimal Power Flow} \label{sec:OPF}

Let $\ccalG = (\ccalV, \ccalE, \ccalW)$ be a graph with a set of $N$ nodes $\ccalV$, a set of edges $\ccalE \subseteq \ccalV \times \ccalV$ and an edge weight function $\ccalW: \ccalE \to \reals_{+}$. We use this graph to model the power grid, where the nodes are the buses and the edge weights $\ccalW(i,j) = w_{ij}$ model the lines, which depend on the impedance $z_{ij}$ (in ohms) between bus $i$ and bus $j$. More specifically, we use the gaussian kernel $w_{ij} = \exp(-k|z_{ij}|^{2})$ where $k$ is a scaling factor. We ignore links whose weight is less than a threshold $\omega$, so that $\ccalE = \{(i,j) \in \ccalV \times \ccalV : w_{ij} > \omega\}$. Additionally, denote by $\bbW \in \reals^{N \times N}$ the adjacency matrix of the network, $[\bbW]_{ij} = w_{ij}$ if $(i,j) \in \ccalE$ and $0$ otherwise. Since the graph is undirected, the matrix $\bbW$ is symmetric.

To describe the state of the power grid, we assign to each node $n$ a vector $\bbx_{n} = [v_{n}, \delta_{n}, p_{n}, q_{n}] \in \reals^{4}$ where $v_{n}$ and $\delta_{n}$ are the voltage magnitude and angle, respectively, and where $p_{n}$ and $q_{n}$ are the active and reactive powers, respectively. We note that the \emph{state} of each node can be accurately measured locally. We can collect these measurements across all nodes and denote them as $\bbv \in \reals^{N}$, $\bbdelta \in \reals^{N}$, $\bbp \in \reals^{N}$ and $\bbq \in \reals^{N}$. Thus, the collection of the states at all nodes becomes a matrix
\begin{equation}
	\bbX = \begin{bmatrix} \bbv & \bbdelta & \bbp & \bbq \end{bmatrix} = \begin{bmatrix} \bbx_{1}^{\Tr} \\ \vdots \\ \bbx_{N}^{\Tr} \end{bmatrix} \in \reals^{N \times 4}.
\end{equation}
Of particular importance are the nodes that are generators, since these are the ones that we are interested in controlling. We proceed to denote by $\ccalV_{G} = \{v_{1},\ldots,v_{M}\} \subset \ccalV$ the set of $M < N$ generators $v_{m} \in \{1,\ldots,N\}$ and define a selection matrix $\bbG \in \{0,1\}^{M \times N}$ that picks out the $M$ generator nodes out of all the $N$ node sin the power grid, i.e. $[\bbG]_{ij} = 1$ if $v_{i} = j$ and $0$ otherwise. We note that $\bbG \bbG^{\Tr} = \bbI_{M}$ is the $M \times M$ identity matrix and that $\bbG^{\Tr} \bbG = \diag(\bbg)$ with $\bbg \in \{0,1\}^{N}$ is the selection vector, i.e. $[\bbg]_{n} = 1$ if $n \in \ccalV_{G}$ and $0$ otherwise. Finally, we note that $\bbG \bbX \in \reals^{M \times 4}$ is the collection of states only at the generator nodes.

The problem of optimal power flow (OPF) consists in determining the power that each generator needs to input into the power grid to satisfy the demand, while minimizing the cost of producing that power at each generator. Let $c_{m}(p_{m})$ be a cost function for generator $m \in \ccalV_{G}$ that depends only on the generating active power $p_{m}$. The active power $p_{n}$ at node $n$ is determined by the voltage magnitudes $\bbv$ and angles $\bbdelta$ at all other nodes and by the topological structure of the power grid $\bbW$. This relationship is given by a nonlinear function $\ccalP$ such that $\bbp = \ccalP(\bbv, \bbdelta; \bbW)$ and is derived from Kirchhoff's current law. Likewise, for the reactive power we have $\bbq = \ccalQ(\bbv, \bbdelta;\bbW)$ for another function $\ccalQ_{n}$. Additionally, there are operational constraints on the power grid, limiting the possible values of active and reactive power, and of voltage magnitude and angles. Let $\bbX^{\min}$ and $\bbX^{\max}$ be the matrices collecting the minimum and maximum value each state entry can take, and denote by $\bbX \preceq \bbY$ the elementwise inequality $[\bbX]_{ij} \leq [\bbY]_{ij}$ for all $i,j$. With all this notation in place, we can write the OPF problem as follows
\begin{align}
\underset{\{p_{m}\}}{\text{minimize}}
	& \sum_{m \in \ccalV_{g}} c_{m}(p_{m}) 
		\label{eqn:objectiveOPF} \\
\text{subject to } 
	& \bbp = \ccalP(\bbv, \bbdelta; \bbW) ,
		\label{eqn:activeOPF} \\
	& \bbq = \ccalQ(\bbv, \bbdelta; \bbW)  ,
		\label{eqn:reactiveOPF} \\
	& \bbX^{\min} \preceq \bbX \preceq \bbX^{\max} .
		\label{eqn:operationalConstraintsOPF}
\end{align}

Solving the OPF problem for DC operation (bias point) is straightforward since the problem becomes convex, as long as the cost function is convex. The cost function is most commonly a second order polynomial of $p_m$. In the DCOPF approximation, functions $\ccalP$ and $\ccalQ$ become linear and the problem can be readily solved. However, when we consider ACOPF dynamics, the problem becomes non-convex due to the computation of the active and reactive power \eqref{eqn:activeOPF}-\eqref{eqn:reactiveOPF}, rendering it computationally costly. In practice, this problem is usually solved using one of five optimization problem solvers: CONOPT, IPOPT, KNITRO, MINOS and SNOPT. However, in general, they are slow to converge for large networks \cite{acopf-computational}. In particular, we consider IPOPT \cite{ipopt} (Interior Point OPTimizer) to provide an optimal solution of the problem (albeit costly) since it was shown to be one of the more robust \cite{acopf-computational}.

The objective of this paper is to \emph{learn} the optimal solution in a decentralized and scalable manner by exploiting a framework known as \emph{imitation learning}. Denote by $\bbp^{\ast}$ the optimal solution obtained by IPOPT, by $\bbPhi(\bbX)$ a (likely, nonlinear) map of the data and by $\ccalL$ some given loss function. Then, in \emph{imitation learning} we want to solve
\begin{equation} \label{eqn:imitationObjective}
    \min_{\bbPhi} \mbE \bigg[ \ccalL\Big( \bbp^{\ast}, \bbPhi(\bbX) \Big)\bigg]
\end{equation}
where the expectation is taken over the (unknown) joint distribution of optimal power $\bbp^{\ast}$ and states $\bbX$. Solving \eqref{eqn:imitationObjective} in its generality is typically intractable \cite{lan2016algorithms}. Therefore, we choose a parametrization of the map $\bbPhi(\bbX) = \bbPhi(\bbX ; \ccalH)$ where $\bbPhi$ now becomes a known family of functions (a chosen model) that is parameterized by $\ccalH$. Likewise, to address the issue of the unknown joint probability distribution between $\bbp^{\ast}$ and $\bbX$ we assume the existence of a dataset $\ccalT = \{(\bbX, \bbp^{\ast})\}$ that can be used to approximate the expectation operator \cite{lan2016algorithms}. With this in place, the problem of imitation learning now boils down to choosing the best set of parameters $\ccalH$ that solves
\begin{equation} \label{eqn:imitationObjectiveParameter}
    \min_{\ccalH} \sum_{(\bbX, \bbp^{\ast})} \ccalL \Big( \bbp^{\ast}, \bbPhi(\bbX; \ccalH) \Big).
\end{equation}

The desirable properties of locality and scalability can be achieved by a careful choice of model $\bbPhi(\bbX;\ccalH)$. In particular, we focus on graph neural networks (GNNs) \cite{Bruna14-DeepSpectralNetworks, Defferrard17-CNNGraphs, Gama19-Architectures} to exploit their stability properties \cite{Gama19-Stability} that guarantee scalability \cite{Eisen19-Wireless, Tolstaya19-Flocking}.


\section{Graph Neural Networks} \label{sec:GNN}

%
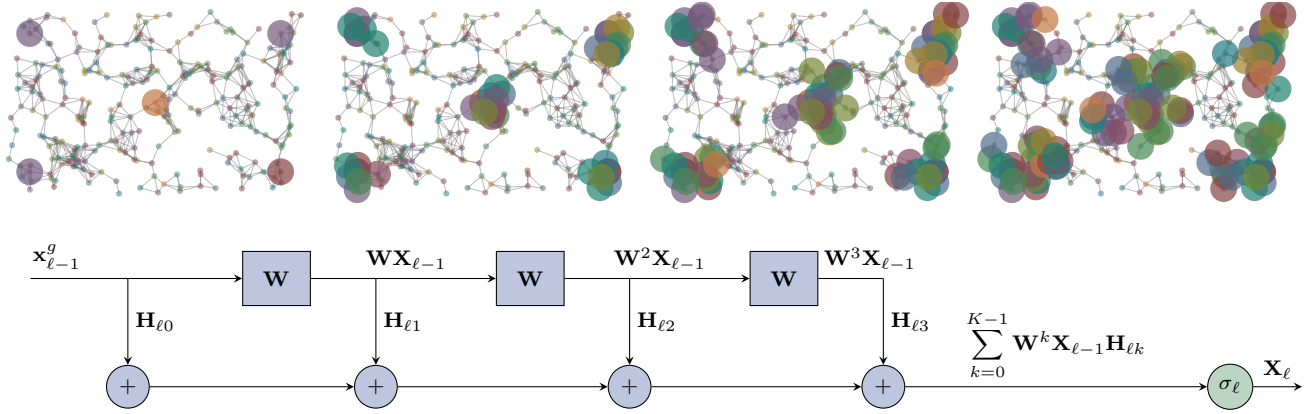
\begin{figure*}[t]
	\centering
	\includegraphics [width = 0.24\linewidth]
	{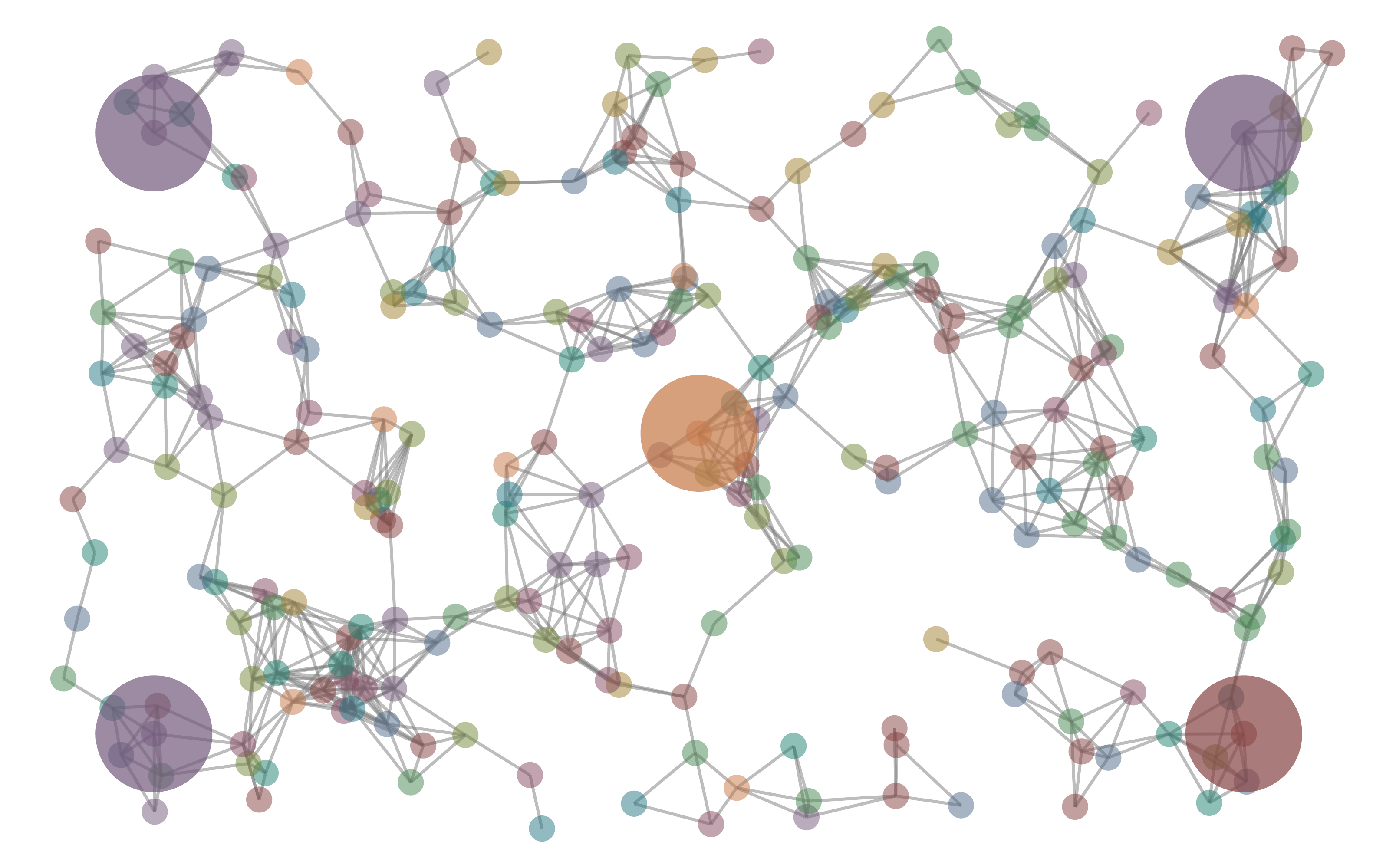}
	\includegraphics [width = 0.24\linewidth]
	{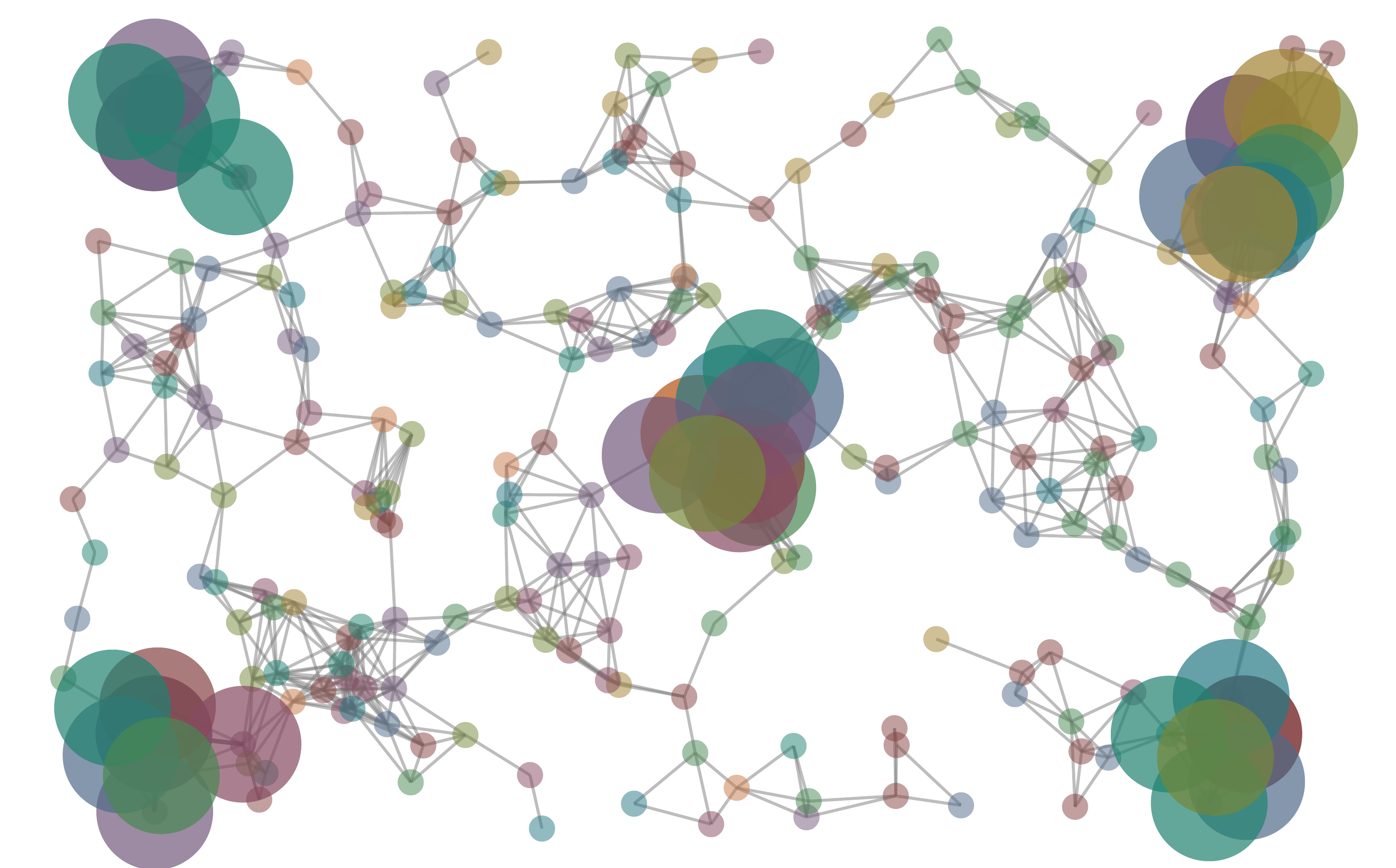}
	\includegraphics [width = 0.24\linewidth]
	{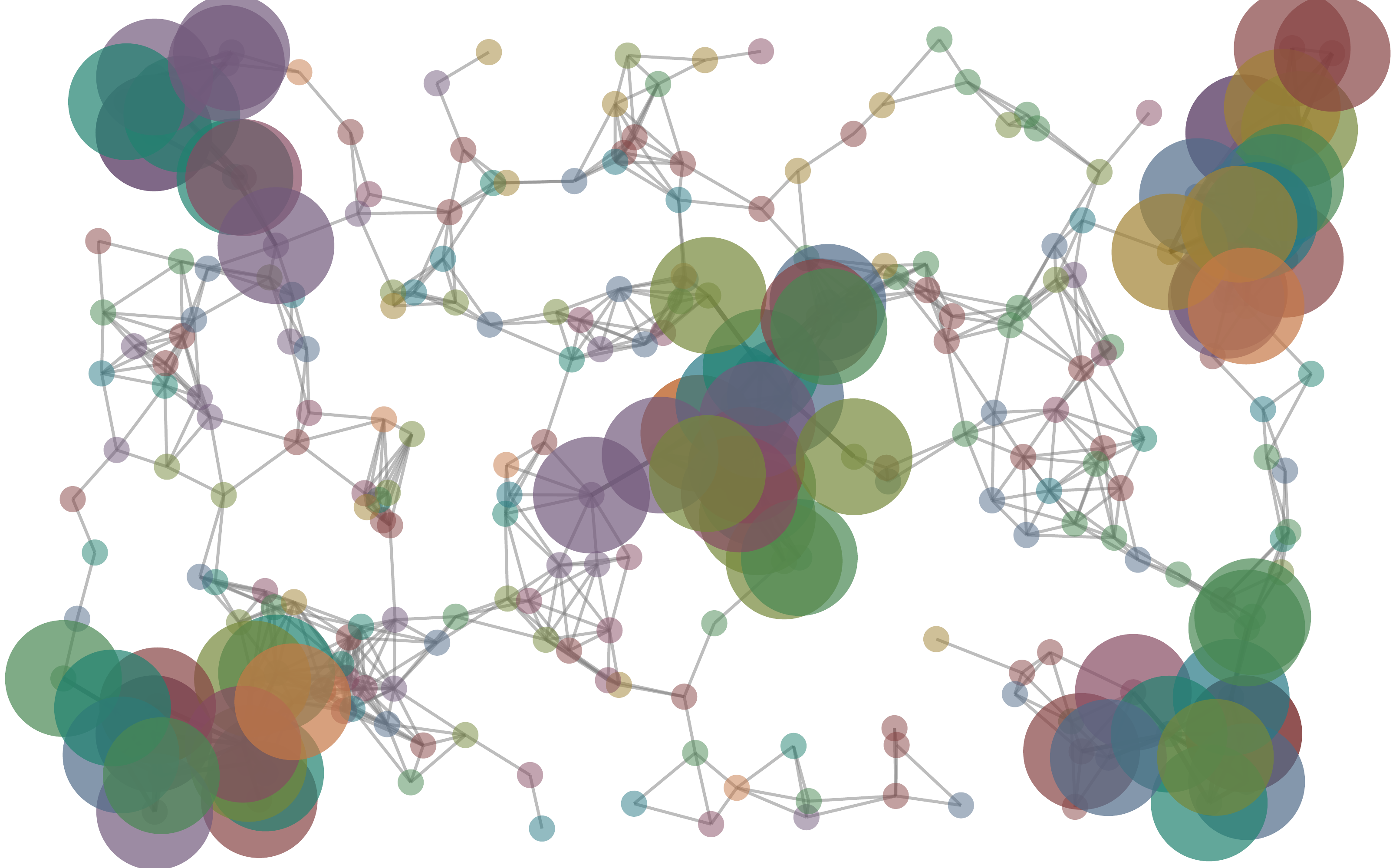}
	\includegraphics [width = 0.24\linewidth]
	{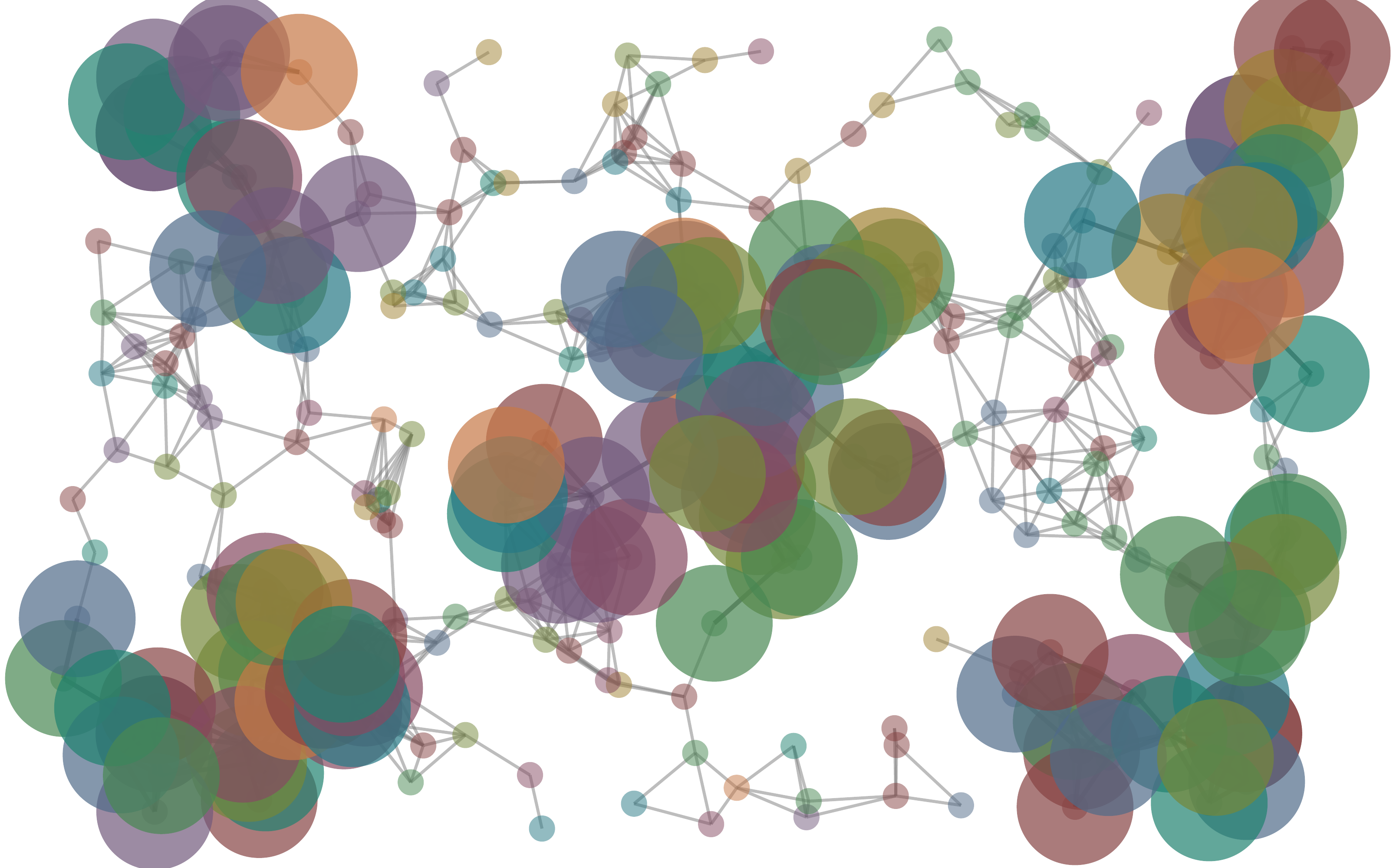} \\ \bigskip

\def \thisplotscale {1.8}
\def \unit {\thisplotscale cm}

\tikzstyle {Phi} = [rectangle, 
                    thin,
                    minimum width = 0.5*\unit, 
                    minimum height = \sumshift*\unit, 
                    anchor = west,
                    draw,
                    fill = pennblue!20]

\tikzstyle {sum} = [circle, 
                    thin,
                    minimum width  = 0.3*\unit, 
                    minimum height = 0.3*\unit, 
                    anchor = center,
                    draw,
                    fill = pennblue!20]

\def \deltax {1.2}
\def \deltay {0.8}
\def \sumshift {0.4}

\begin{tikzpicture}[x = 1*\unit, y = 1*\unit] \footnotesize
    
\node (origin) [] {};
\path (origin) ++ (0.2*\deltax, 0) node (first) [] {};
    
\path (first) ++ (1.35*\deltax, 0) node (0) [Phi] {$\bbW$};
\path (0)     ++ (1.35*\deltax, 0) node (1) [Phi] {$\bbW$};
\path (1)     ++ (1.35*\deltax, 0) node (2) [Phi] {$\bbW$};

\path (2.east) ++ (1.0*\sumshift*\deltax, 0) node [anchor=west] (last) [] {};

\path (first.east) ++ (1.5*\sumshift*\deltax, -\deltay) node (sum0) [sum] {$+$};
\path (0.east) ++ (\sumshift*\deltax, -\deltay) node (sum1) [sum] {$+$};
\path (1.east) ++ (\sumshift*\deltax, -\deltay) node (sum2) [sum] {$+$};
\path (2.east) ++ (\sumshift*\deltax, -\deltay) node (sum3) [sum] {$+$};

\path[-stealth] (first) edge [very near start, above] node {$\bbx_{\ell-1}^{g}$}               (0);	
\path[-stealth] (0)     edge [above] node {$\ \bbW\bbX_{\ell-1}$}     (1);	
\path[-stealth] (1)     edge [above] node {$\ \bbW^{2}\bbX_{\ell-1}$} (2);	
\path[-]        (2)     edge [above, near end] node {$\ \bbW^{3}\bbX_{\ell-1}$} (sum3 |- last);

\path[-stealth, draw] (sum0 |- first) -- (sum0) node [midway, right] {$\bbH_{\ell 0}$};	
\path[-stealth, draw] (sum1 |- 0)     -- (sum1) node [midway, right] {$\bbH_{\ell 1}$};	
\path[-stealth, draw] (sum2 |- 1)     -- (sum2) node [midway, right] {$\bbH_{\ell 2}$};	
\path[-stealth, draw] (sum3 |- 2)     -- (sum3) node [midway, right] {$\bbH_{\ell 3}$};

\path[-stealth, draw] (sum0) -- (sum1);	
\path[-stealth, draw] (sum1) -- (sum2);	
\path[-stealth, draw] (sum2) -- (sum3);	

\path (sum3.east) ++ (2.0 * \deltax, 0) node (nonlinearity) [sum, fill=penngreen!20] {$\sigma_{\ell}$};
\path[-stealth,draw] (sum3) -- (nonlinearity) node [midway, above] {$\displaystyle \sum_{k=0}^{K-1} \bbW^{k} \bbX_{\ell-1} \bbH_{\ell k}$};

\path[-stealth] (nonlinearity.east) edge [midway, above] node {$\bbX_{\ell}$} ++ (0.3*\deltax, 0);



\end{tikzpicture}
	\caption{Graph neural networks. Every node takes its data value $\bbX_{\ell-1}$ and weighs it by $\bbH_{\ell 0}$ (first graph). Then, all the nodes exchange information with their one-hop neighbors to build $\bbW\bbX_{\ell-1}$, and weigh the result by $\bbH_{\ell 1}$ (second graph). Next, they exchange their values of $\bbW\bbX_{\ell-1}$ again to build $\bbW^{2}\bbX_{\ell-1}$ and weigh it by $\bbH_{\ell 2}$ (third graph). This procedure continues for $K$ steps until all $\bbS^{k} \bbX_{\ell-1}\bbH_{\ell k}$ have been computed for $k=0,\ldots,K-1$, and added up to obtain the output of the graph convolution operation \eqref{eqn:graphConvolution}. Then, the nonlinearity $\sigma_{\ell}$ applied to compute $\bbX_{\ell}$. To avoid cluttering, this operation is illustrated on only $5$ nodes. In each case, the corresponding neighbors accessed by successive relays of information are indicated by the colored disks.} 
	\label{fig:selection}
\end{figure*}
%

In the context of graph signal processing (GSP) the state $\bbX  \in \reals^{N \times F}$ of the nodes can be seen as a \emph{graph signal} \cite{Sandryhaila13-DSPG, Shuman13-SPG, Ortega18-GSP}. Essentially, row $n$ corresponds to the $F=4$ measurements or \emph{features} obtained at node $n$. We can relate the graph signal $\bbX$ to the underlying network support by multiplying by $\bbW$ to the left
\begin{equation} \label{eqn:graphShift}
    \bbY = \bbW \bbX.
\end{equation}
The output $\bbY \in \reals^{N \times F}$ is another graph signal that is a \emph{shifted} version of the input $\bbX$. To see this, note that now the $f$th feature at node $i$, $[\bbY]_{if} = y_{i}^{f}$, is a linear combination of the $f$th feature values $[\bbX]_{jf} = x_{j}^{f}$ at nodes $j$ that are in the the one-hop neighborhood of $i$, $\ccalN_{i} = \{j \in \ccalV : (j,i) \in \ccalE\}$. More specifically,
\begin{equation} \label{eqn:graphShiftDetailed}
     y_{i}^{f} = \sum_{j = 1}^{n} [\bbW]_{ij} [\bbX]_{jf} = \sum_{j \in \ccalN_{i}} w_{ij} x_{j}^{f}.
\end{equation}
where the second equation follows due to the sparsity of the adjacency matrix $\bbW$. Based on \eqref{eqn:graphShiftDetailed}, operation \eqref{eqn:graphShift} can be interpreted as a diffusion or a \emph{shift} of the signal across the graph, where each state gets updated by a linear combination of the neighboring states. While in this paper we focus on the use of the adjacency matrix $\bbW$, any other matrix that respects the sparsity of the graph (i.e. $[\bbW]_{ij} = 0$ if $(i,j) \notin \ccalE$) works. Matrices that satisfy this constraint are called \emph{graph shift operators} (GSOs) and other examples of widespread use in the GSP literature include the Laplacian matrix, the random walk Laplacian and several normalized versions of these \cite{Ortega18-GSP}.

The concept of \emph{shift} \eqref{eqn:graphShift} is used as the basic building block for the operation of \emph{graph convolution}. In analogy with its discrete-time counterpart, a \emph{graph convolution} is defined as a linear combination of shifted version of the input signal
\begin{equation} \label{eqn:graphConvolution}
    \bbY = \sum_{k=0}^{K-1} \bbW^{k} \bbX \bbH_{k}
\end{equation}
with $\bbH_{k} \in \reals^{F \times G}$ the matrix of coefficients for $k=0,\ldots,K-1$. The output $\bbY \in \reals^{N \times G}$ is another graph signal with $G$ features on each node. As we analyzed in \eqref{eqn:graphShiftDetailed}, the operation $\bbW \bbX$ computes a linear combination of neighboring values. Likewise, repeated application of $\bbW$ computes a linear combination of values located farther away, i.e. $\bbW^{k} \bbX$ collects the feature values at nodes in the $k$-hop neighborhood. The value of $\bbW^{k} \bbX = \bbW (\bbW^{k-1} \bbX)$ can be computed locally by $k$ repeated exchanges with the one-hop neighborhood. We note that multiplication by $\bbH_{k}$ on the right does not affect the locality of the graph convolution \eqref{eqn:graphConvolution}, since $\bbH_{k}$ acts by mixing the features local to each node. That is, it takes the $F$ input features, and mixes them linearly to obtain $G$ new features. To draw further analogies with filtering, we observe that the output $\bbY$ of the graph convolution \eqref{eqn:graphConvolution} is the result of applying a bank of $FG$ linear shift-invariant graph filters \cite{Segarra17-Linear}.

A graph neural network (GNN) is a nonlinear map $\bbPhi(\bbX; \ccalH, \bbW)$ that is applied to the input $\bbX$ and takes into account the underlying graph $\bbW$. It consists of a cascade of $L$ layers, each of them applying a graph convolution \eqref{eqn:graphConvolution} followed by a pointwise nonlinearity $\sigma_{\ell}$ (see Fig.~\ref{fig:selection} for an illustration)
\begin{equation} \label{eqn:GNN}
    \bbX_{\ell} = \sigma_{\ell} \Bigg[ \sum_{k=0}^{K_{\ell}-1} \bbW^{k} \bbX_{\ell-1} \bbH_{\ell k} \Bigg]
\end{equation}
for $\ell=1,\ldots,L$, where $\bbX_{0}=\bbX$ the input signal and $\bbH_{\ell k} \in \reals^{F_{\ell-1} \times F_{\ell}}$ are the coefficients of the graph convolution \eqref{eqn:graphConvolution} at layer $\ell$  \cite{Bruna14-DeepSpectralNetworks, Defferrard17-CNNGraphs, Gama19-Architectures}. The state $\bbX_{\ell} \in \reals^{N \times F_{\ell}}$ at each layer $\ell$ is a graph signal with $F_{\ell}$ features and we consider the state at the last layer $\bbX_{L}$ to be the output of the GNN $\bbPhi(\bbX;\ccalH, \bbW) = \bbX_{L}$.

The computation of the state $\bbX_{\ell}$ in each of the $\ell$ layers can be carried out entirely in a local fashion, by means of repeated exchanges with one-hop neighbors. Also, the number of filter taps in $\bbH_{\ell k}$ is $F_{\ell-1}F_{\ell}$, independent of the size of the network, and thus, the GNN \eqref{eqn:GNN} is a scalable architecture \cite{Gama19-Architectures}. This justifies the consideration of the GNN \eqref{eqn:GNN} as the model of choice in \eqref{eqn:imitationObjectiveParameter}, $\bbPhi(\bbX; \ccalH) = \bbPhi(\bbX; \ccalH, \bbW)$ with parameters $\ccalH = \{\bbH_{\ell k}, k=0,\ldots,K_{\ell-1}, \ell = 1,\ldots,L\}$ totaling $|\ccalH| = \sum_{\ell=1}^{L} F_{\ell-1}F_{\ell}K_{\ell}$, independent of the size $N$ of the network. Furthermore, GNNs exhibit the properties of permutation equivariance and stability to graph perturbations \cite{Gama19-Stability}. The first one, allows the GNN to learn from fewer datapoints by exploiting the topological symmetries of the network, while the second one allows the network to have a good performance when used on different networks than the one it was trained in, as long as these networks are similar.

In summary, in this paper we propose using the GNN \eqref{eqn:GNN} as a local and scalable model $\bbPhi(\bbX; \ccalH, \bbW)$ that we use to \emph{imitate} the optimal solution $\bbp^{\ast}$. We find the best model parameters $\ccalH$ by solving \eqref{eqn:imitationObjectiveParameter} over a given dataset $\ccalT$.


\section{Numerical Experiments} \label{sec:sims}

We construct two datasets based on the IEEE-30 and IEEE-118 power system test cases \cite{matpower}. Each dataset sample consists of a given load, where $\bbp^L \in \reals^N, \bbq^L \in \reals^N$ are the load components at each node, a given sub-optimal actual state of the power grid $\bbX,$ and the optimal power generated at each node, $\bbp^{\ast}$ (computed by means of IPOPT). The total power at each node is the difference between the generated power, $\bbp^G \in \reals^N$ and $\bbq^G \in \reals^N$, and the load at that node, $\bbp^L, \bbq^L$. Certainly, only generators are capable of generating power, while all other buses just consume their power load. These are related to the OPF equations \eqref{eqn:activeOPF} and \eqref{eqn:reactiveOPF} as follows
\begin{align}
	p_m &= [\bbp^G]_m - [\bbp^L]_m, & m \in \ccalV_{G} \label{eq:pgl} \\
	q_m &= [\bbq^G]_m - [\bbq^L]_m, &  m \in \ccalV_{G} \label{eq:qgl} \\
	p_{n} &= - [\bbp^L]_{n}, & n \in \ccalV \backslash \ccalV_{G} \label{eq:pl} \\
	q_{n} &= - [\bbq^L]_{n}. & n \in \ccalV \backslash \ccalV_{G} \label{eq:ql}
\end{align}

%
\begin{figure}
	\centering
	\includegraphics[width=0.9\linewidth]{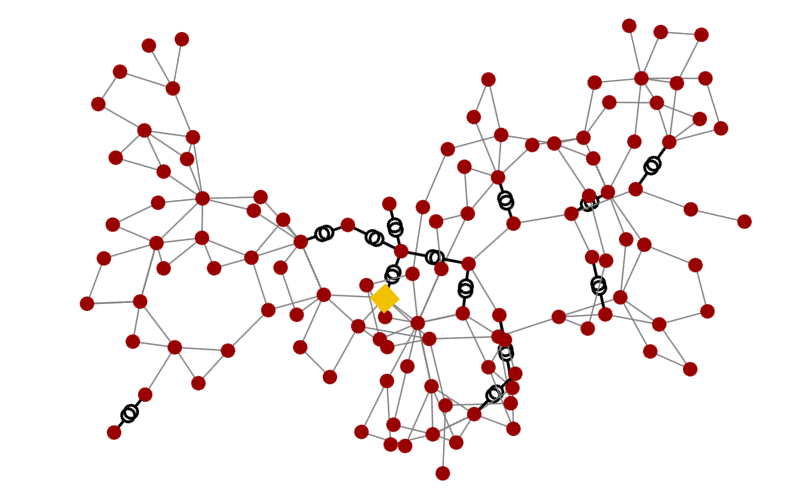}
	\caption{Diagram of IEEE-118 power system test case. Red circles are buses, the yellow square is an external grid connection and intersecting circles are transformers.\vspace{-0.3cm}}
	\label{fig:118}
\end{figure}
%

The test cases provide reference load, $\bbp^L_{\text{ref}}$ and $\bbq^L_{\text{ref}}$ \cite{matpower}. Each given load is obtained synthetically as a sample from the uniform distribution around the reference, following the same methodology used in \cite{guha}
\begin{align}
\bbp^L & \sim \text{Uniform}(0.9\ \bbp^L_{\text{ref}}, 1.1 \ \bbp^L_{\text{ref}}) \label{eq:loadsP} \\
\bbq^L & \sim \text{Uniform}(0.9\ \bbq^L_{\text{ref}}, 1.1\ \bbq^L_{\text{ref}}) \label{eq:loadsQ}
\end{align}
The sub-optimal state $\bbX$ is obtained by using Pandapower \cite{pandapower} to find the steady state of the system, also known as a power flow computation. We can interpret this steady state as the state of the power grid before we need to adjust it to accommodate for the new values of $\bbp^{L}$ and $\bbq^{L}$. Meanwhile, the steady-state generator power is set by solving a DCOPF to find a sub-optimal generator configuration. DCOPF has a low computational cost and is therefore often used to warm start ACOPF. The generated load is also used as an input to ACOPF, which we solve using IPOPT \cite{ipopt} in Pandapower to obtain the optimal generator output $\bbp^{\ast}$. We discard samples for which IPOPT fails to converge. Therefore, for the IEEE-30 and IEEE-118 test cases, we generate $8,016$ and $13,129$ samples, respectively. These are split $80\%$ for training and $20\%$ for testing. We use this dataset to compare four architectures: global GNN, local GNN, global MLP and local MLP, which are described next.

The global GNN architecture consists of two graph convolutional layers [cf. \eqref{eqn:GNN}] and one readout layer. The first convolutional layer has $K=4$ filter taps and outputs $F_{1} = 128$ features. The second layer also has $K=4$ taps and outputs $F_{2}=64$ features. These are followed by a last readout layer consisting of a fully connected layer with input size $64N$ and output size $M$. We note that the inclusion of the fully connected layer makes the global GNN architecture nonlocal, since it arbitrarily mixes the features at all nodes. Also, the number of parameters of the entire architecture depends on the size of the graph. Therefore, we also test a local GNN architecture, where we keep the first two convolutional layers, but we change the readout layer. More specifically, this new readout layer computes a linear combination of the features available locally at each node to output a single scalar to represent the estimated power. This is equivalent to setting $K=1$ in the last layer of \eqref{eqn:GNN}, $\bbX_L = \sigma_L \left[ \bbX_{L-1} \bbH_L \right]$, where $\bbH_L \in \reals^{F_{L-1} \times 1}$ represents the linear combination of input features with $\bbX_L \in \reals^{N}$. Finally, we select only the outputs at the generator nodes $\bbG\bbX_L \in \reals^{M}$. As we can see, in the local GNN, all operations involved can be carried out by repeated exchanges with one-hop neighbors and do not involve centralized knowledge of the state of the entire power grid.

As a baseline, we compare the GNN to the global MLP used in \cite{guha}. The MLP has two hidden layers with $128$ and $64$ features, respectively, thus having $128N$ and $64N$ hidden units. We keep the readout layer to be a fully connected layer mapping $64N$ into the $M$ estimated power at the generators. Since this solution is also centralized, we use another baseline consisting of a local MLP that only performs operations on the state $\bbX$ of each node. Specifically, we set $K=1$ in all layers of the GNN [cf. \eqref{eqn:GNN}], so that it becomes becomes $\bbX_{\ell} = \sigma_{\ell} \left[ \bbX_{\ell-1} \bbH_{\ell} \right]$, $\bbH_{\ell} \in \reals^{F_{\ell-1} \times F_{\ell}}$. As with the local GNN, we consider $\bbG\bbX_L$ to be the output. We note that this last architecture, albeit local, it completely ignores the underlying graph structure of the data, since it does not involve any kind of neighbor exchanges.

We train the aforementioned architectures over the IEEE-30 and IEEE-118 training sets by solving \eqref{eqn:imitationObjectiveParameter} for an MSE loss function. The optimal $\bbp^{\ast}$ is given by the solution of IPOPT, and the parametrization $\bbPhi$ under consideration is each one of the four architectures, i.e. $\bbPhi(\bbX; \ccalH, \bbW)$ for the global and local GNNs, and $\bbPhi(\bbX; \ccalH)$ for the global and local MLPs (which, as becomes explicit now, do not take into account the underlying graph structure). We use an ADAM optimizer with learning rate $0.001$ and forgetting factors $\beta_1 = 0.9$ and $\beta_2 = 0.999$. We train for a $100$ epochs with a batch size of $128$. We evaluate the trained architectures on the corresponding test set by using a performance measure given by the relative root mean squared error $\sqrt{\| \bbp^{\ast} - \bbPhi(\bbX; \ccalH) \| / \| \bbp^{\ast}\|}$ averaged over all values of $(\bbX, \bbp^{\ast})$ in the test set. Results are shown in Table~\ref{tab:results}.

\begingroup
\setlength{\tabcolsep}{10pt} 
\renewcommand{\arraystretch}{1.5} 
\begin{table}
	\centering
    \caption{The RMSE for each architecture and dataset.}
	\begin{tabular}{|c|c|c|}
		\hline 
		Architecture & IEEE 30 & IEEE 118 \\ 
		\hline 
		Global GNN & 0.061 & 0.00306 \\ 
		\hline 
		Global MLP & 0.090 & 0.00958 \\ 
		\hline 
		Local GNN & 0.139 & 0.03038 \\ 
		\hline 
		Local MLP & 0.161 & 0.35932 \\ 
		\hline 
	\end{tabular}
    \label{tab:results}
\end{table}
\endgroup

We note that in both test cases, the GNNs outperform the MLPs. In the smaller network IEEE-30, the relative improvement of the Global GNN over the Global MLP is of $47\%$, while for the Local GNN with respect to the local MLP is of $15\%$. Likewise, for the larger network IEEE-118, the Global GNN relatively improves $213\%$ over the Global MLP, while the Local GNN exhibits a relative improvement of $1082\%$ over the Local MLP. This shows that local solutions that exploit the underlying network structure outperform other similar neural network parametrizations that ignore it.

Finally, we note that obtaining the IPOPT solution $\bbp^{\ast}$ based on a given state $\bbX$ takes $2.16\text{s}$ on the IEEE-30 network and $18\text{s}$ on the IEEE-118 one, representing a $8$-fold increase in the computational time. One the other hand, the Global GNN and the Local GNN take $48 \mu\text{s}$ and $49\mu\text{s}$, respectively, on the IEEE-30 dataset, and $50\mu \text{s}$ and $56\mu \text{s}$, respectively, on the IEEE-118 dataset. This amounts to $1.04$ and $1.14$ times more computational time, respectively. This shows that, not only the use of GNNs is over $10^{5}$ faster, but that they also exhibit much better scalability to larger networks.



\vspace{-0.2cm}

\section{Conclusions} \label{sec:conclusions}

Solving the OPF problem is central to power grid operation. OPF is a non-convex problem whose exact solution can be computed by means of IPOPT (interior point optimizer). However, this solution is costly and does not scale to large networks. In this paper, we proposed the use of GNNs to approximate the OPF solution. GNNs are scalable and local information processing architectures that exploit the network structure of the data. We train GNNs by taking a given network state as input, and using the corresponding output to approximate the optimal IPOPT solution, following the imitation learning framework. We run experiments on the IEEE-30 and IEEE-118 test cases, showing that local solutions that adequately exploit the underlying grid structure outperform other comparable methods.



\bibliographystyle{IEEEbib}
\bibliography{myIEEEabrv,biblioOPF}

\end{document}